# Simulation studies on laser pulse stability for Dalian Coherent Light Source[*]

DENG Hai-Xiao(邓海啸) [1)]   ZHANG Meng(张猛) [2)]   GU Duan(谷端)   LIU Bo(刘波)
GU Qiang(顾强)   WANG Dong(王东)
Shanghai Institute of Applied Physics, the Chinese Academy of Sciences, Shanghai 201800, China

**Abstract** Dalian Coherent Light Source will use a 300MeV LINAC to produce fully coherent photon pulses in the wavelength range between 150-50nm by high gain harmonic generation free electron laser (FEL) scheme. To generate stable FEL pulses, stringent tolerance budget is required for the LINAC output parameters, such as the mean beam energy stability, electron bunch arrival time jitter, peak current variation and the transverse beam position offset. In order to provide guidance for the design of the Dalian Coherent Light Source, in this paper, the sensitivity of FEL pulse energy fluctuation to various error sources of the electron bunch was performed using intensive start-to-end FEL simulations.

## 1. Introduction

Free electron lasers (FEL) hold great prospects as high power photon source with tunable wavelength. In order to satisfy the dramatically growing demands within the material, biological and chemical sciences, many FEL user facilities have been constructed and being proposed worldwide, from the extreme ultraviolet (EUV) to hard x-ray spectral region [1-5]. Currently, FEL community is on a stage to more sophisticated schemes, e.g., in pursuit of compact size [6-9], flexible polarization [10-15] and fully coherence [16-22]. Dalian Coherent Light Source (DCLS) is a FEL user facility based on the principle of single-pass, high-gain harmonic generation [16], which is located in northeast of China. With the state-of-the-art techniques of optical parametric amplification (OPA) seed laser and flexible gap undulator, DCLS is dedicated at EUV spectral regime of 150-50nm with pulse energy exceeds 100μJ.

At present, it is believed that the scientific application for DCLS will firstly focus on the Physical Chemistry with time-resolved pump-probe experiments and EUV absorption spectroscopy techniques. The time domain experiments are mainly related to the FEL pulse duration, e.g., less than 100fs, which can be easily achieved by utilizing an ultra-short seed pulse for DCLS. However, another important and inevitable topic associated with both the time and spectral domain experiments for DCLS is FEL laser pulse stability, i.e., shot-to-shot fluctuation, which will be strongly sensitive to numerous error sources of the electron beam and the see laser time jitter. According to the FEL physical design and corresponding tolerance budgets illustrated in Section 2, this paper numerically investigates on the laser pulse stability for DCLS. Section 3 mainly concentrate on the jitter results of the LINAC output parameters, which will serve as the input conditions for the start-to-end FEL simulations in Section 4. This paper is concluded with the final remarks in Section 5.

## 2. FEL design and tolerance budgets

DCLS is the first high gain FEL user facility in China. The layout of DCLS (seen in Fig. 1) includes a seed laser system, injector system, linear accelerator and undulator system. The injector and LINAC provide electron beams with the energy of 300MeV, bunch charge up to 500pC, peak current up to 300A, and the normalized transverse emittance better than 1μm-rad. The undulator system consists of 1 segment of modulator and 4 segments of radiator, where the period length is 50mm and 30mm, respectively. The Ti-Sa laser system can feed seed pulse with energy of 100μJ, wavelength range of 250-400nm by the OPA technique, and the pulse duration can be switched between 1ps and 130fs, thus to generate FEL pulse with different temporal properties.

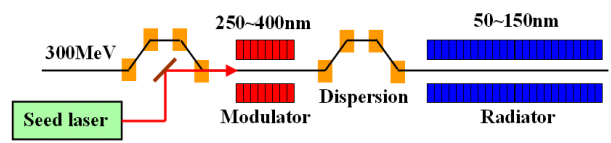

Fig. 1. Layout of Dalian Coherent Light Source.

DCLS generates FEL pulses with continuously tunable frequency, which is truly useful to the specific FEL users. In order to maximum the FEL pulse energy, with a fixed conversion efficiency from electron to light, the electron beam power should be as large as possible for interested wavelength. Thus, for different radiation wavelength for DCLS, one should tune the wavelength of OPA seed laser and the undulator gaps simultaneously to satisfy FEL resonance while keep the electron beam energy at maximum 300MeV. Fig. 2 shows the peak power of the radiation pulse at different wavelength for DCLS, in which the well-known empirical analytical formulas [23] and the well benchmarked FEL code, GENESIS [24] is used. The results show that, under the conditions of the bunch charge of 500pC, seed laser pulse of 1ps (FWHM)

---

Received 18 April 2013
* Supported by National Natural Science Foundation of China (21127902, 11175240 & 11205234) and the Knowledge Innovation Program of Chinese Academy of Sciences.
1) E-mail: denghaixiao@sinap.ac.cn   2) E-mail: zhangmeng@sinap.ac.cn

and beam energy of 300MeV, the output pulse energy should be 100μJ level and the photon number per pulse can be up to $10^{13}$ order. It is worth to stress that DCLS performances are optimized at the spectral region larger than 100nm, since the users are especially interested in the efficiency excitation of hydrogen atom and methyl radical [25]. Moreover, the undulator tapering technique could further augment the DCLS output by a factor of two with no difficulty.

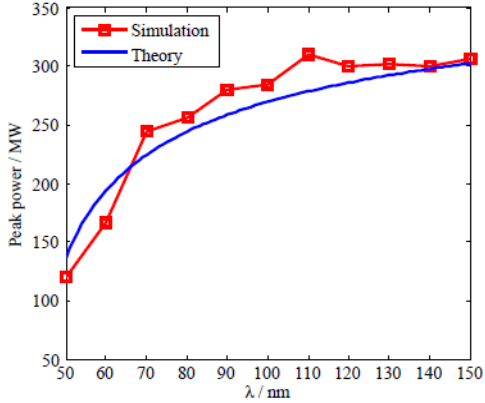

Fig. 2. Analytical and numerical results of DCLS output peak power at different radiation wavelength.

To estimate the sensitivity of DCLS output, extensive time-dependent GENESIS simulations were performed in which various electron beam quantities and the seed laser timing were varied independently around their ideal values. As the tolerance budgets summarized in Table 1, the beam energy, bunch charge, normalized emittance, peak current and slice beam energy spread requirements are from the specification of 100μJ output pulse energy at 50 nm, i.e., the wavelength with the largest sensitivity.

Table 1. Main tolerance budgets of DCLS

| Parameter | Specification |
| --- | --- |
| RMS peak current jitter | < 10% |
| RMS normalized emittance (μm-rad) | < 1 |
| RMS slice energy spread (keV) | < 20 |
| RMS beam position jitter (μm) | < 10 |
| RMS beam energy jitter | < 0.1% |
| RMS beam arrival time jitter (fs) | < 150 |
| RMS seed laser timing jitter (fs) | < 100 |

In DCLS user operation, shot-to-shot repeatability of the output photon number should be as good as possible. Ideally, the shot-to-shot fluctuation in normalized pulse energy is expected to be less than 5% for the nonlinear phenomena experiments, which seems unlikely with the presently expected accelerator and injector performance. However, many of experiments can tolerate fluctuation as high as 25% by recording the shot-to-shot pulse photon number for post-processing [26]. The sensitivity scans show that the initial beam energy jitter plays a crucial role in the DCLS output stability, especially at 50nm wavelength, where the FEL pierce parameter [27] is only 0.14%. Then a 10μm transverse beam position jitter at the undulator entrance obviously requires a beam based alignment procedure for LINAC [28]. Finally, 1ps electron beam pulse and 1ps seed pulse seems much relaxed the tolerance requirement on timing between the electron beam and the seed laser.

## 3. Jitter sensitivity from LINAC

Photo-cathode RF injector and one bunch compressor LINAC is used to generate 300A peak current electron beam for DCLS. Before the bunch compressor chicane, two S-band accelerating structures (L1-L2) generate the required energy chirp for bunch compression, and four S-band accelerating structures (L3-L6) after is used to accelerate the electron beam to 300MeV, while removing the correlated energy spread. FEL operations foresee stringent requirements for the stability of the LINAC output parameters, such as the beam arrival time, peak current, and average beam energy. In order to understand the jitter sensitivity of these parameters to various error sources along the LINAC, an elaborate study using particle tracking codes has been performed.

The beam sensitivity of LINAC is investigated by summing the uncorrelated random effects for DCLS, i.e., 250fs injector timing jitter, 5% bunch charge jitter, 0.1% RF voltage jitter, 0.1° RF phase jitter and 0.02% bunch compressor dispersion jitter, which leads a 0.09% beam energy jitter, 5.33% peak current jitter and 137fs beam arrival time jitter for current LINAC design of DCLS, as shown in Fig. 3. According to the tracking results, the uppermost source of the mean beam energy jitter is the accelerating structures L1-L2 where the linear energy chirp is generated, the beam current jitter mainly comes from the injector, and the beam arrival time jitter is almost evenly distributed along the LINAC.

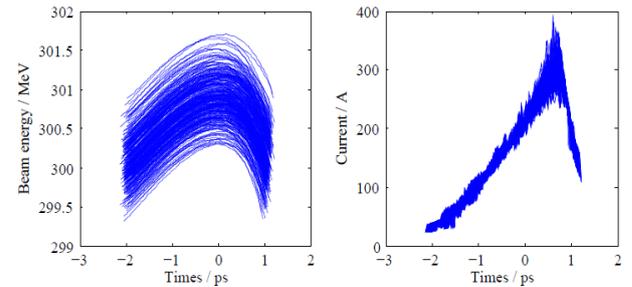

Fig. 3. LINAC tracking results of 300 shots start-to-end simulations, without X-band cavity.

The electron orbit in LINAC is strongly related on the quadrupole magnet offsets. Without proper correlation, the maximum beam transverse error could be up to 1.5mm, which is far from the requirement for FEL lasing. Therefore, a beam based correction [28] is used firstly to minimize the quadrupole offsets with respect to the beam orbit. Then one random orbit with maximum offset is chosen for beam transverse jitter calculation because of the implicit influence. A 0.1% RMS jitter is applied to all the power supplies of the magnets which have effects on the beam orbit, e.g., bending magnets, correctors and quadrupoles, no more than 10μm beam jitter is modeled in both horizontal and vertical axis.

X-band structure is widely utilized for linearizing the bunch compression process in large-scale hard x-ray FEL. And here we consider LINAC beam dynamics for DCLS with X-band cavity. Considering its small beam energy and compact machine size, the electron beam should be chirped more for a flatter top current profile for DCLS, as shown obviously in Fig. 4. On the basis of these settings and tracking results, one can easily conclude that the mean beam energy jitter for the X-band case is slightly more stringent than the nominal case, while the sliced energy spread and the sliced transverse emittance are almost similar in both cases.

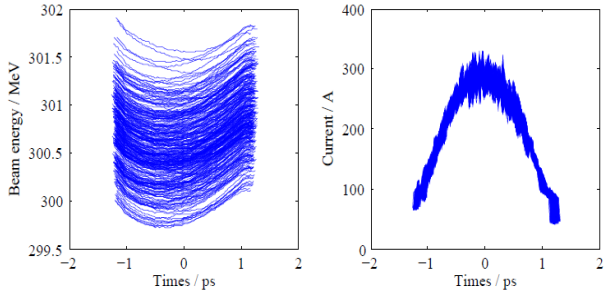

Fig. 4. LINAC tracking results of 300 shots start-to-end simulations, with X-band linearization.

## 4. Start-to-end simulations on FEL pulse stability

By importing those 300 shots LINAC tracking results to the DCLS undulator system, we obtain the start-to-end statistical information on FEL performance under two tolerance sets, as shown in Table 2.

Table 2. Start-to-end laser pulse stability of DCLS

| X-band | OFF | ON |
|---|---|---|
| LINAC RF amplitude jitter | 0.1% | 0.1% |
| LINAC RF phase jitter (degree) | 0.1 | 0.1 |
| LINAC power supply jitter | 0.1% | 0.1% |
| Beam energy jitter | 0.09% | 1.20% |
| Beam peak current jitter | 5.9% | 4.0% |
| Beam transverse position jitter (μm) | 10/5 | 10/10 |
| Beam arrival time jitter (fs) | 140 | 138 |
| Seed laser timing jitter (fs) | 100 | 100 |
| 50nm photon number fluctuation | 26% | 42% |
| 100nm photon number fluctuation | 16% | 15% |
| 150nm photon number fluctuation | 10% | 12% |

FEL simulation setup used here has been optimized in terms of undulator lattice and seeding parameters in order to maximize the output peak power extracted from an ideal bunch. The setup for the start-to-end simulations is the following: the 5MW seed laser with a Gaussian profile (1ps FWHM in longitudinal and 0.5mm RMS in transverse) and a 100fs RMS timing jitter, the modulator tuned at 250nm, the radiator tuned at 50nm and the dispersive section with $R_{56} = 0.18$mm. As illustrated in Fig. 5, the mean pulse energy and the standard deviation of 50nm FEL is about 60μJ, 26% and 72μJ, 42% for the cases with X-band OFF and ON, respectively.

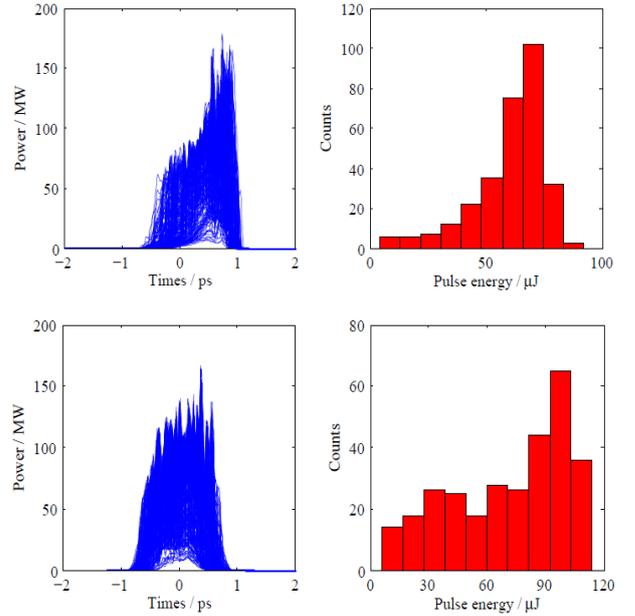

Fig. 5. Start-to-end results of 50nm pulse distributions from 300 shots. Top: without X-band cavity, Low: with X-band linearization.

50nm wavelength of DCLS shows the most stringent requirement on the quality and tolerance of the electron beam. When we compared with other error sources, e.g., laser-beam timing and beam transverse position jitter, the mean beam energy jitter is almost close to the pierce parameter of 50nm FEL. According to the start-to-end LINAC tracking results, the X-band cavity is helpful for compressing more electron particles to the beam core with a larger energy chirp before the bunch compressor, and thus leads a 0.12% RMS beam energy jitter, which is larger than the 0.09% beam energy jitter associated with the X-band structure OFF. Therefore, the X-band cavity enhances the radiation power and purifies the radiation spectrum of 50nm FEL while reduces its pulse stability.

Moreover, when one considers the undulator resonant equation, the FEL wavelength jitter should be two times that associated to the mean beam energy jitter. However, this is not true in DCLS where the radiation wavelength is mainly defined by the seed laser and partially depends on the undulator resonant wavelength. The DCLS output spectra results are in agreement with predictions, and the obtained fluctuation of the central wavelength is much lower when compared to the beam energy jitter, and thus not affecting the FEL performance.

In order to verify the conclusions on 50nm start-to-end simulations, which indicates the mean beam energy jitter as the most limiting factor for achieving a good output stability, the DCLS output pulse stabilities of 100nm and 150nm cases are also characterized, in which the FEL pierce parameters is several times larger than the electron beam energy jitter and the FEL saturates in 2 segments of radiator. Fig. 6 and 7 shows the corresponding output pulse distribution and the pulse energy statistics from the start-to-end simulation results. The average pulse energy and the standard deviation of the 100nm FEL distribution is about 248μJ, 16% and 186μJ, 15% for the cases with and without X-band cavity. Meanwhile, the average pulse energy and the deviation of the 150nm FEL is about 468μJ, 10% and 296μJ, 12% for the cases with and without the X-band structure, respectively. Here, the larger pulse energy for the case without X-band structure is mainly benefited from the beam current spike which exceeds the nominal 300A, as shown in Fig. 3.

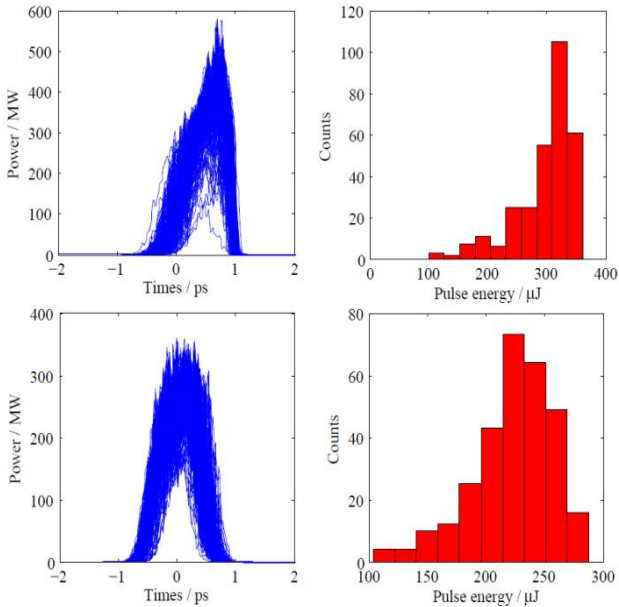

Fig. 6. Start-to-end results of 100nm pulse distributions from 300 shots. Top: without X-band cavity, Low: with X-band linearization.

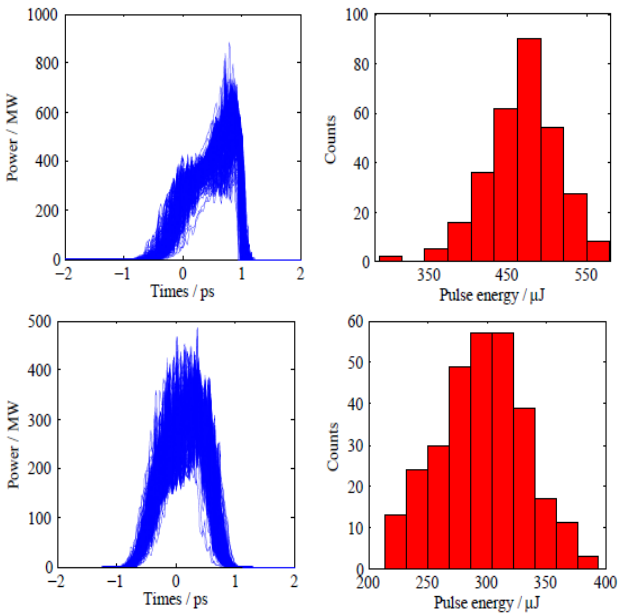

Fig. 7. Start-to-end results of 150nm pulse distributions from 300 shots. Top: without X-band cavity, Low: with X-band linearization.

## 5. Conclusions

Dalian Coherent Light Source is expected to generate fully coherent laser pulses by seeded FEL scheme in the wavelength range between 150-50nm with pulse photon number of $10^{13}$ order. In this paper, with the help of intensive start-to-end FEL simulations, DCLS laser pulse stability sensitivity to various error sources, e.g., magnet power supply instability, RF jitter and timing jitter was investigated. FEL pulse stability performance is mainly affected by the electron beam energy jitter for DCLS. For the current design of DCLS machine without X-band, the RMS variation of laser pulse energy is about 26% for the 50nm radiation wavelength and becomes much better in the long wavelength regime.

For comparative purpose, FEL pulse stability with the X-band cavity was also studied in this paper. In order to compress more electrons to the beam core with a larger energy chirp before the bunch compressor, the X-band cavity leads 0.12% beam energy jitter, which contributes to the FEL stability degradation at short-wavelength for DCLS. However it is worth stressing that, the X-band linearization is helpful for suppression of the coherent synchrotron radiation effects in the bunch compression and removal of the correlated beam energy chirp after the bunch compression. Thus in further studies, the X-band structure demonstrates a better distribution of emittance, energy spread and beam current, and thus a much more pure radiation spectrum for DCLS.

*The authors would like to thank Tong Zhang, Chao Feng, Lie Feng, Taihe Lan, Xingtao Wang and Zhentang Zhao for helpful discussions.*